# Fermi's Paradox Is a Daunting Problem – Under Whatever Label

Milan M. Ćirković[1]
*Astronomical Observatory of Belgrade, Volgina 7,*
*11000 Belgrade, Serbia*

## 1. Introduction

Gray (2015) argued that Fermi's paradox (FP) is a misnomer, and it is not a valid paradox. Gray also speculated that the argument was misattributed to Fermi, whose lunchtime remarks did not pertain to the existence of extraterrestrial intelligence, but to the feasibility of interstellar travel. Instead, the paradox is ascribed to Hart and Tipler, and it is further suggested that the paradox is not a "real" problem or research subject and should not be used in debates about SETI projects.

The arguments given are unpersuasive, ahistorical, and, in at least one instance, clearly hinge on literalistic and uncharitable reading of evidence. Instead, I argue the following three points: (i) Contrary to Gray's assertion, the historical issue of naming of ideas or concepts is completely divorced from their epistemic status. (ii) FP is easily and smoothly generalized into the "Great Silence" paradox, so it makes no sense either theoretically or empirically to separate the two. (iii) In sharp contrast to the main implication of Gray's paper, FP has become more aggravated lately due to advances in astrobiology.

Research that deals with FP has greatly expanded in recent years on both a theoretical and observational stage (Davies 2010, 2012; Vukotić and Ćirković 2012; Barlow 2013; Hair and Hedman 2013; Davies and Wagner 2013; Armstrong and Sandberg 2013; Lampton 2013; Cartin 2014; Nunn, Guy, and Bell 2014; Wright et al. 2014; Spivey 2015; Griffith et al. 2015; Zackrisson et al. 2015 – to give just a few post-2010 examples). Not all of these papers have been authored by astronomers or astrobiologists; the multidisciplinary nature of the whole effort is seen in titles such as "The Fermi Paradox and Coronary Artery Disease" in one of the world's most prestigious medical journals (Gottlieb and Lima 2014), and the subject matter is discussed even by personalities such as the (in)famous NSA whistleblower Edward Snowden.[2] Multiple editions of books such as that of Webb (2015), a renowned physicist and author, testify on both research and public interest in the topic.

If we were to accept Gray's arguments, we would have not only denigrated this effort (and much more, since the bibliography of Fermi's paradox would easily amount to several thousand references), we would also have done it for no cognitive gain. Fortunately enough, as I will show, there is no need to accept them. In sharp contrast to Gray's view, FP is not an obstacle, but a great *research opportunity* in astrobiology, SETI, and future studies. Therefore, Gray is correct that FP should not be construed as inimical to SETI studies; in contrast, it boosts original, productive, and proactive work in the field. Speculation that it is possible that FP played a role in lawmakers decisions (in a single country) who obstructed SETI does not, and can not, impact cognitive and epistemic value of FP-related research.

---

[1] Also at Future of Humanity Institute, Faculty of Philosophy, University of Oxford, Suite 8, Littlegate House, 16/17 St Ebbe's Street, Oxford, OX1 1PT, UK. E-mail: mcirkovic@aob.rs.

[2] As reported at http://www.theguardian.com/us-news/2015/sep/19/edward-snowden-aliens-encryption-neil-degrasse-tyson-podcast, last accessed March 1, 2016.

## 2. Smell of a rose and misnomers

*What's in a name?* asked Juliet Capulet. Does that which we call a rose smell differently in the context of astrobiology? Should we abandon researching FP if we were to gain evidence that our framing of the problem is not how Enrico Fermi read it? The fact that a concept is called by the name of X need not mean that X is more than peripherally relevant to the concept itself. Using Fermi's lunchtime comments as the only true and authoritative formulation of the problem is as appropriate as the hypothetical usage of Lou Gehrig's sport biography in a study of pathophysiology of Lou Gehrig's disease. At the very best it can give us a *particular* instance, which one could then confuse with the *general* issue at one's own peril. Fermi's remarks are part of the *history* of astrobiology and SETI, not of astrobiology and SETI themselves.

Consider the Copernican Revolution, about which we certainly do not learn today from the book of Copernicus. There is no particular importance to circular orbits in the real world of planetary science, as there was in Copernicus' writings, and the Sun is certainly not the mystical source of all light and good as it was described in *De Revolutionibus*. The Copernican Revolution has become a widely important *historical* phenomenon in years and decades and even centuries after the death of Copernicus himself; so to associate it for any serious purpose with the Canonical Text would be a massive fallacy. Similarly, one can be an excellent relativist without ever reading a word of Einstein's original writings on the subject or a distinguished evolutionist without reading Darwin. The same applies to FP, since Fermi did not research the topic in any way. Further, his casual lunchtime remarks should be considered even *less* a canonical view of the problem than might be the book of Copernicus (who was doing diligent research on planetary motions) as a textbook on the Copernican Revolution. The greatest strength of science lies in its capacity for generalization; if an argument could easily be made more general – hence stronger – it is only scientifically and intellectually honest to face the most general version, *irrespective of its historical genesis and naming*.

Gray asserts that "[s]ome people may feel that the so-called Fermi paradox is a sleeping dog that should be left to lie, because it is established... most people would agree that clearly mistaken and misleading terminology should be corrected." Contrary to what Gray implies, naming of concepts is not part of the research activity in a field, but a part of history, sociology, and administration of science. Cognitive value of FP stays the same if we decide to call it Joe's problem or Cindy's conjecture or Singh's puzzle – how can it be otherwise? Long ago Plato warned that "the name is not a thing of permanence, and that nothing prevents the things now called round from being called straight, and the straight things round; for those who make changes and call things by opposite names, nothing will be less permanent (than a name)."[3]

Many – if not most –concepts in astronomy are misnomers; in an ancient science that has always possessed so much dynamism, it is perhaps to be expected. Let me offer just a few examples most people would emphatically *not* agree to as correct. Apparent, absolute, visual and other MAGNITUDES of celestial bodies, especially stars, are not really magnitudes (the word meaning bigness or size); all stars were point-like sources until very recently. There is no Ocean of Storms or Sea of Tranquillity, since LUNAR MARIA are not seas, in spite of being thus *named*. PLANETARY NEBULAE have nothing whatsoever to do with planets; NOVA (and especially SUPERNOVA) is not, contrary to its Latin meaning since the time of Tycho, a new star, but an old one. The usage of METALS (and derivatives like METALLICITY) to denote carbon or oxygen or sulphur is likely to make any physical chemist cry, since they exhibit no metallic properties whatsoever. EARLY- and LATE-TYPE GALAXIES form neither chronological nor evolutionary sequence and therefore are neither early nor late.

And, of course, cosmology is rife with such misnomers. The celebrated HUBBLE CONSTANT is *variable* in most cosmological models, including the realistic one. VOIDS are not really devoid of matter, and the

---

EPOCH OF RECOMBINATION did not, in contrast to laboratory plasmas, mean *re*combining of electrons and nuclei, since they had never been together in the first place. A Gray-like project of excising misnomers would have to insist on renaming it the *epoch of combination* – not an appealing proposal to most cosmologists. The ANTHROPIC PRINCIPLE has nothing to do with man (ἄνθρωπος), but deals instead with the observation-selection effects common to any observer. Speaking about paradoxes, OLBERS' PARADOX can hardly bear scrutiny, since Digges, Kepler, Halley, and Cheseaux have all had better claim on it than H. W. Olbers.

So, shall we embark upon the comprehensive project of excising misnomers? Or shall we peacefully come to terms with *history* as an inseparable part of any human endeavour, including science, and devote time and resources to solving *real* problems? Literalism has never been particularly successful in any field, and it is illusory to think it would be different in astrobiology and SETI studies. Even if Gray were correct about the original meaning, re-writing of history is illusory – in 65+ years of *history* of FP there has been so much fruitful research activity (as well as so much popular science and pop-culture) that excising would make as much sense as trying to excise planetary nebulae. There is no connection between the issues of adequate naming of entities and concepts, which is a *historical* and *administrative* issue, and the epistemic and cognitive status of those entities and concepts. Would we worry less about what FP can tell us about, for instance, the future of humanity (Baum 2010), if we decide to call it *Hart's problem* instead?

## 3. FP is the Great Silence

The locution „Fermi's paradox" should not – if we wish to have a substantive discussion of ideas and not just a scholastic discussion about words – be used literally for Fermi's lunchtime remarks, *whose exact content is anyway unknown with certainty*, but as synonymous with the more general and precise Great Silence paradox (Brin 1983; Ćirković 2009). The Great Silence paradox has nothing in particular to do with exploration or conquest, and even less with any form of human psychology or history; it does not necessarily have anything to do with the feasibility of interstellar travel either. Instead, the Great Silence paradox has to do with the general *detectability* of other intelligent species.

**The Great Silence:** *The lack of any detectable activities or manifestations or traces of extraterrestrial civilizations in our past light cone is incompatible with the multiplicity of such civilizations and conventional assumptions about their capacities.*

*Where is everyone?* should not be construed as pertaining to physical visits only, but to anything and everything detectable due to intentional activity. Why is it *too damn quiet* in the universe around us (Kent 2011)? Physical visits are a special case of the more general set of traces and manifestations of intelligent beings. That much was clear to Brin 33 years ago. Although Brin's comprehensive study has been cited in Gray's paper, it appears only in the beginning, and its pertinent views are not discussed. In particular, Brin justifies why the classical FP is *subsumed* into the more general Great Silence problem, and there is neither need nor gain to separate the two.

Each important concept in the history of science first appeared in rudimentary and often *wrong* form. Subsequent work involves making concepts more general, more precise, taxonomically ordered, etc. Who really cares that Copernicus was wrong about circular orbits or that Darwin held nebulous ideas about the mechanism of inheritance? Meaningful opposition to the original theory of evolution by means of natural selection drew on the problem of dilution of selected characters allowed by the pre-Mendelian understanding of heredity ("blending inheritance"). Such anti-Darwinian arguments were *perfectly legitimate*; the fact that it ultimately turned out that Darwin was right and Fleeming Jenkin wrong (Gould 2002) does not impact the criticism of Darwin's original theory.

Gray's paper gives the impression that it is obvious that Fermi only doubted the feasibility of interstellar travel, not the existence of extraterrestrial intelligent beings. While the historical record is ambiguous on this point,[4] it is irrelevant to the general problem. Unfeasibility of interstellar travel is only *one* of a number of possible solutions to FP, and so why commit the logical error of mistaking a part for the whole? If Fermi indeed perceived just that one way of addressing the problem, so much the worse for him (although it seems implausible). There is no need for a modern SETI researcher to follow the same road, no *more* need than it was for a planetary astronomer to follow Copernicus in insisting on circular planetary orbits. A circle is a special case among ellipses; unfeasibility of interstellar travel is a special case within the family of solutions to FP. That Copernicus was wrong about circularity does not impact his contribution in asking the right kind of question; that Fermi was (probably) wrong about interstellar travel[5] does not impact his contribution in asking the right kind of question.

## 4. More serious than ever

In its Great Silence form, FP has become aggravated in recent years. The crucial contribution (and an excellent example of an important astrobiological result neglected in SETI circles) is the work of Lineweaver (2001) on the age distribution of terrestrial planets in the Milky Way. His calculations show that Earth-like planets began forming more than 9.2 Ga ago, and that their median age is $t_{med} = (6.4 \pm 0.9) \times 10^9$ yrs, which is significantly greater than the age of the solar system. A large majority of habitable planets, including the oldest among them, are much older than the Earth. The application of Copernican principle would then immediately suggest that the stage of the biospheres and even the stage of evolution of advanced technological civilizations must be, on the average, older than the stage we see on Earth by almost 2 Gyr. This difference is large in comparison with the Galaxy timescales with regard to crossing/colonizing/astro-engineering/filling with intentional messages. This is in sharp contrast not only with the absence of visitors, but also with the absence of any manifestations and artefacts of older Galactic civilizations.

This constitutes a paradox (not necessarily a purely logical paradox, since many paradoxes in the physical sciences are not really logical paradoxes – see Maxwell's demon, Olbers' paradox). And the tension has been aggravated over recent decades due to several independent lines of scientific and technological advances. One of these is the discovery of more than 3000 extrasolar planets, to date, with more discoveries occurring on an almost daily basis (http://exoplanet.eu/). Some of the other items that generally undermine the naive Copernican picture and aggravate FP include confirmation of the *rapid* origination of life on early Earth; an improved understanding of molecular biology and biochemistry that has led to heightened confidence in the theories of naturalistic abiogenesis; exponential growth of the technological civilization on Earth, as especially manifested through Moore's Law and other advances in information technologies; an improved understanding of the *feasibility* of interstellar travel in both a classical sense and in the more efficient form of sending inscribed matter packages; theoretical grounding for various astroengineering projects detectable over interstellar distances; an improved understanding of the extragalactic universe that has engendered a wealth of information about other galaxies, many of them similar to the Milky Way (and, it should be noted, not a single civilization of Type 3 or high 2.x has been found, in spite of the huge volume of space surveyed). The recent burst of research activity cited in the introduction is well motivated and timely.

---

[4] Both Teller's and York's quotations given by Gray actually refer to low spatiotemporal density of extraterrestrials (something to be explained!), with York adding the important point about possible short lifetimes of intelligent communities as the main reason for not engaging in interstellar travel. The implication, clearly, is that if lifetimes were longer, interstellar travel would have made perfect sense – nothing obviously unfeasible about interstellar travel itself.

[5] E.g., Long (2012).

## 5. Conclusions

While critical discussions of issues related to FP are certainly welcome, a revisionist account such as Gray's does not seem to contribute to further understanding. Whatever name we choose for what is at least informally known as Fermi's paradox, this will remain a challenging problem – indeed, the central one – for SETI studies and the sector of astrobiology that deals with high-complexity life. Amidst exciting new observational and theoretical work, the room for semantic discussions on the origin and proper naming of concepts is smaller by the day. Indeed, as suggested by a great historian of science long ago, the true verification of SETI studies as proper science depends on its capacity to leave such scholasticism behind (Dick 1996). Nowhere is this truer than in the studies of Fermi's paradox.